\documentstyle[aps,pra]{revtex} 
\begin{document} 
\title{Quantum dynamics of evaporatively cooled Bose-Einstein Condensates}
\author{P. D. Drummond and J. F. Corney}
\address{Department of Physics, The University of Queensland, Australia 4072}
\date{\today}
\maketitle

\begin{abstract}
We report on dynamical simulations
of Bose-Einstein condensation via evaporative
cooling in an atomic trap. The results show evidence for spontaneous
vortex formation and quantum dynamics in small traps.

\end{abstract}


Evaporative cooling has been successfully used to produce 
Bose-Einstein condensates (BEC)
inside magneto-optic traps with neutral atoms\cite{BEC}.  A number of
questions arise as to the quantum state
 that is achieved, since this involves
both the dynamics of the cooling process and the applicability of the
ergodic hypothesis.
Atom-atom
interactions have a strong influence on the cooling
process and the final state in these experiments.
Quantum fluctuations are important in determining atom laser coherence 
properties\cite{MewAndKurDurTowKet97}, especially since
the experimental systems do not have as large a particle
number as traditional condensed matter experiments. However,
there is no guarantee that a canonical ensemble will result
from evaporative cooling, as the
observations are made in a transient, non-equilibrium phase.
Thus,
conventional canonical methods may not be applicable
to these experiments.

In this paper, we report the use of phase-space methods for direct quantum dynamical
calculations of the cooling and formation of Bose-Einstein condensates on a three-dimensional lattice.
The results are restricted as yet to small condensates, due to the 
large numbers of modes involved. The computational results  are very similar
to those  observed experimentally. In particular,
we find
quantum evaporative cooling, followed by a clear transition to a condensate.
This is strongly influenced by non-classical features of the quantum dynamics. 
The calculations indicate additional structure, which we interpret as 
spontaneous
formation of vortices - a process of much wider interest in other fields of physics
\cite{Zurek}. These appear to originate in the 
residual orbital
angular momentum of the trapped atoms, which was neglected in previous
studies, and would provide a significant test of the present theory.

Earlier calculations of cooling dynamics have usually treated
the cooling process either
classically\cite{Theor,Ber97}, or have used various additional
assumptions about the quantum states involved. This leads to the
question of how to handle the transition to the final quantum
dominated condensate, which is often assumed to be a canonical
ensemble at a temperature estimated from the classical theory.
The final ensemble behaviour is then usually calculated from
the mean-field Gross-Pitaevskii 
equations\cite{BijSto97},
 although some attempts have been made to go beyond this
\cite{ZieShu97}, including treatment
of the kinetics of condensation\cite{Kinetics,Kinetics2}
based on a master equation.
However,
 small atom traps are neither in the
thermodynamic limit, nor necessarily in a steady-state. 
A first principles theory is really needed, to provide
a benchmark for comparisons of these previous approximations,
like the quantum Monte Carlo (QMT)  theories\cite{Ceperley} in equilibrium systems.

In our calculations, we include $3\times 10^4$ relevant modes (which is
a very conservative estimate), with up to $1.0\times 10^4$ atoms
present. 
The quantum state-vector therefore has over $10^{10000}$ components.
One possible approach in principle is to use quantum
number state calculations in the time-domain.
Any direct calculation which includes all the relevant modes of the
trapped atoms - up to the energy scales required for evaporating
atoms to escape - is easily seen to be an enormous computational
problem. 

A more practical technique is to utilize phase-space methods that have 
already proved successful in 
laser theory. These techniques can handle large numbers
of particles, but can also systematically treat
departures from classical behaviour, including
boson interactions. Generalized phase-space representations were used to 
correctly predict quadrature squeezed quantum soliton dynamics in optical 
fibers\cite{CDRS87}, which are described by nearly identical quantum
equations to those 
used in atom-atom interaction studies. 
The
coherent-state (positive-$P$) phase-space equations
 are exactly equivalent to the relevant quantum equations -
 provided phase-space boundary terms\cite{CDRS87} vanish. They have the 
advantage that they are computationally tractable for
the large Hilbert spaces typical of BEC experiments. Techniques of this sort can
provide a first step towards extending QMT methods\cite{Ceperley} into the time-domain. 

The model that we use includes the usual non-relativistic
Hamiltonian for neutral atoms
in a trap $V({\bf x})$, interacting via a potential $U({\bf x})$, together
with absorbing reservoirs $\hat R ({\bf x})$, in $d=2$ or $d=3$
dimensions:

\begin{eqnarray}
\hat{H} = \int d^d{\bf x}\biggl[
{\hbar^2 \over 2m} \nabla \hat \Psi^\dagger({\bf x})\nabla \hat \Psi({\bf 
x}) + V({\bf x})\hat\Psi^{\dagger}({\bf x})\hat\Psi({\bf x})
 +
\hat\Psi^\dagger({\bf x})\hat R({\bf x}) + \hat\Psi({\bf x}) \hat R^\dagger 
({\bf x})\biggr. 
+ {1 \over 2}\biggl. \int d^d {\bf y} U({\bf x} -{\bf y})
\hat \Psi^\dagger ({\bf x}) \hat \Psi^\dagger ({\bf y})
\hat \Psi({\bf y}) \hat \Psi({\bf x}) \biggr] \, .
\end{eqnarray}

Here $\hat R({\bf x})$
 represents a localized  absorber that removes the neutral atoms 
- for
example, via collisions with foreign atoms, or 
at the location of the `RF-scalpel' resonance, which is used
to cause evaporative cooling\cite{BEC}. We 
expand 
$\hat \Psi$ using free-field modes with a momentum cut-off $k_{max}$.
Provided that $k_{max} << a_0^{-1}$, where $a_0$ is the 
S-wave scattering length,
$U({\bf x} - {\bf y})$ can be replaced by the renormalized pseudo-potential
$u\delta^d ({\bf x}-{\bf y})$, where $u=4\pi a_0 \hbar^2 /m$ in three 
dimensions.  In two dimensions, $u$ is defined similarly but with a factor $\xi_0$
in the denominator, which corresponds to the effective spatial extent of the condensate in the third  direction. This factor is of the order of the lattice spacing in the simulation, and is chosen to be equal to $x_0$, the scaling length.

The resulting quantum time-evolution for the density matrix
$\hat \rho$ can be solved  by expanding into a coherent-state
basis, and then (provided phase-space boundary terms vanish)
transforming to an equivalent set of 
equations in the positive-$P$ representation.
 The phase-space equations in the BEC case  can be expressed as 
 two coupled complex partial stochastic differential equations
of the form:

\begin{eqnarray}
 i\hbar{\partial \psi_j\over\partial t}&=&\biggr[{-\hbar^2\over 2m}
 \nabla^2+u\psi_j\psi_{3-j}^*+V({\bf x})-{i\hbar\over 
2}\Gamma({\bf x})+\sqrt{i\hbar u}
 \xi_j(t,{\bf x})\biggr] \psi_j
\, ,
 \label{+P}
 \end{eqnarray}
 where $j=1,2$ and where the stochastic fields $\psi_j$ are the coherent state amplitudes of a 
non-diagonal coherent
 state projector, $|\psi_1\rangle\langle \psi_2|/\langle 
\psi_2|\psi_1\rangle $.  These equations can be
 readily simulated numerically\cite{WD97} in one, two or three transverse 
dimensions, 
with either attractive or repulsive potentials.  The form of the potentials was
chosen to be 
\begin{eqnarray}
V({\bf x},t) = (1-\alpha t) V_{max} \Sigma_{j=1}^d [\sin(\pi  x_j /L_j)]^{2}
\, ,
\label{V}
\end{eqnarray}
where $\alpha$ is typically the inverse of the total simulation time.The potential height was swept downwards linearly in time, thus
successively removing cooler and cooler subpopulations of atoms.  
The absorption rate $\Gamma({\bf x})$ was chosen as
\begin{eqnarray}
\Gamma({\bf x}) = \Gamma_{max} \Sigma_{j=1}^d [\sin(\pi x_j /L_j)]^{50}
\, .
\label{Gamma}
\end{eqnarray}
Here $L_j$ is the trap 
width in the j-th direction, such that $-L_j/2 \le x_j \le L_j/2$.  The sinusoidal shape of the potential and absorption was chosen so that the trap would be harmonic near the center of the trap, and smoothly approach a maximum near the edge.  Thus hot atoms 
are absorbed when they reach regions
 of large $\Gamma({\bf x})$, located near the trap edges.

A useful feature of Eq (\ref{+P}) is that in the 
deterministic limit,
this corresponds precisely to the well-known Gross-Pitaevskii 
equations, with the addition of a 
  coefficient $\Gamma({\bf x})$ for
 the absorption of atoms by the reservoirs. Quantum effects 
come from the terms 
$\xi_j$, which  are
  real Gaussian stochastic fields, with correlations: 
\begin{equation}
\langle\xi_1(s,{\bf x})\xi_2(t,{\bf y})\rangle=\delta_{ij}
\delta(s-t)\delta^d
({\bf x}-{\bf y}) \, .
\end{equation}
 The quantum correlations that can be calculated include 
 $n({\bf k}) = \langle\psi_1({\bf k})\psi_2^*({\bf k})\rangle$, which 
gives the
 observed momentum distribution.
 
 The results of the simulations depend critically on the exact
 parameters chosen, just as one would expect from the known
 sensitivity of the experiments to the precise experimental conditions. 
 In practical
 computations, it is necessary to consider rather small traps.
 This is 
because the numerical lattice spacing used to sample
 the stochastic fields in x-space must be of order $\Delta x =1/k_{max}$,
 where $k_{max}$ is the largest ordinary momentum considered in the
 problem. However, the  value of the corresponding kinetic
 energy, $E_K = (\hbar k_{max})^2/2m$, must be large enough to allow
 energetic
 atoms to escape over the potential barrier of the trap, otherwise no
 cooling can take place. This sets an upper-bound on the lattice
 spacing, and hence on the maximum trap size - which depends on the 
number of lattice points
 that can be computed.
 
The available lattice sizes used here were $32^d$ points,
depending on the dimensionality $d$. With this limit, and parameter values 
similar to
those used in the experiments, the available trap sizes that can be treated
are of the order of micron dimensions. These are  smaller than those
used  currently, although traps of this type are quite feasible.
 The other possibility within
the constraints is to use a trap which is of larger dimensions but lower in
potential height.  For this type of trap which was simulated here,
the width was $L_j= 10 \mu m $, with a potential height of 
$V_{max}/k_B = 1.9 \times 10^{-7} K$, and an initial temperature of 
$T_0 = 2.4 \times 10^{-7} K$.
 
For physical reasons, a further limitation is that the initial density must be
 such that  $\langle n(k)\rangle \le 1$ - otherwise the starting 
point would already
 have a Bose-Einstein condensation. This places a limit on the number of 
atoms
 which can simulated,
if we assume an initially non-condensed grand canonical ensemble
of (approximately) non-interacting atoms.
  There were initially around $500$ atoms in the 
 two dimensional simulations reported here,  and $10,000$ in 
the three dimensional
 case. These corresponded to atomic densities
of $n_0 = 5.0\times 10^{12} /m^2$ and $n_0 = 1.0\times 10^{19}  /m^3$ 
respectively.

For the small trap parameters used in the simulations, the effect of the stochastic terms on the dynamics is very large.  In fact the quantum fluctuations that these stochastic terms introduce are much larger than the initial thermal fluctuations, such that the initial features of the distribution do not persist.  This means  that the choice of the initial state of the system is not critical, and also that in order to determine properties of the final quantum ground state of the system, the stochastic terms are vital.  For comparison, we investigated the effect of removing the quantum noise
terms, so that the simulations were simply of the Gross-Pitaevskii equation,
with initial conditions corresponding to a thermal state.  For our parameters, these situations did not show  strong Bose
condensation effects, in contrast to the fully quantum-mechanical simulations.
This demonstrates the highly non-classical nature of
the early stages of Bose-condensation, in which spontaneous transitions
to the lowest energy states clearly play an important role.

For the simulations shown in the figures, 
$a_0 = 0.6 nm$ and the mass, corresponding to rubidium, is $m = 1.5\times 10^{-25} kg$. 
These parameters correspond to relatively weakly
interacting atoms, in order to reduce the sampling error - which increased rapidly with longer times and larger coupling constants.
No large phase-space excursions were observed with these parameters.
All results
are plotted in normalized units, with space scaled by $x_0 = 0.76\mu m$,
and time scaled by $t_0 = 0.79 ms$. 
The time-step was typically $t_0/2500$,
with all calculations being repeated at half the time-step (and noise
sampled from the same process with twice the resolution\cite{WD97}), to check
numerical convergence.
The boundary absorption term was set to $\Gamma_{max}  \simeq 10^3 /s$.

In momentum space, the
final atom density for individual trajectories in both two and three dimensions is quite narrow and tall,
with a width corresponding to a temperature well below the critical temperature
for BEC.  
The peak final momentum state population is much greater than 
one (and greater than the initial conditions).  This is more pronounced
in the three dimensional case than in two dimensions, showing that the evaporative
cooling process is more efficient with the extra degree of freedom and the
greater number of atoms that are present. 

 As is usual in quantum mechanics, only the ensemble
averages of the simulations have an operational meaning. Thus,
while individual stochastic realizations have a definite coherent phase,
 these phases are different
for distinct stochastic realizations -- 
the ensemble average has no absolute phase information.
  The average evolution of a two dimensional condensate is shown in Fig. (1); in this case,
  the condensate is only weakly occupied:  
  
Since the condensate does not have to form in the ground-state, the Bose-condensed peaks that occur at different momentum values in single runs are averaged out in the overall ensemble.  A more useful indication of condensation is given by the following measure of phase-space confinement:
\begin{equation}
Q = \frac{\int d^3k \langle\psi_1({\bf k})\psi_1({\bf k})\psi_2^*({\bf k})\psi_2^*({\bf k})\rangle}{\left(\int d^3k\langle\psi_1({\bf k})\psi_2^*({\bf k})\rangle\right)^2x_0^3}.
\end{equation} 
This higher order correlation function is the quantum analogue of the participation ratio defined by Hall\cite{Hall99}.  Figure (2) shows the evolution of $Q$ calculated from 15 runs of the three dimensional simulation.  The sharp rise near $t=100$ is a strong indication of condensation occurring at this point.

For the finite-size condensates in atom traps, just as the final ground state is not expected to be precisely the zero momentum eigenstate, so too 
such condensates are not 
constrained to fall into the $J = 0$ angular momentum eigenstate.  Both the initial and the escaping atoms have
an arbitrary angular momentum.  We can estimate that the variance in angular
momentum will scale approximately as $\langle\hat J^2\rangle \propto N$,
from central limit theorem arguments.
Thus, we can expect that each trapped condensate should have angular 
momentum, unless constrained by the trap geometry.
The angular momentum can be carried either by quasi-particles or vortices, 
although
a volume-filling $j$-th order  vortex has $J= Nj$, and therefore
cannot form spontaneously in the thermodynamic limit of large N.
For small condensates, a  $j=\pm 1$ vortex may be quite
likely.  Several 
authors\cite{VORTEX} 
have considered how such vortex states may form through
stirring or rotating a condensate, and the stability
of vortices has been explored\cite{Stability}. Here we 
consider the possibility of vortices
forming spontaneously in the condensate through the process of 
evaporative cooling, without external intervention. 

The presence of vortex states 
can be detected quantitatively by
transforming the spatial lattice into a lattice which uses the
angular momentum eigenstates as a basis set.  
The two-dimensional
results, which are presented here, are obtained by integrating the spatial
profile over orthogonal modes with corresponding field operators 
$\hat\Psi_{jn}$.
The angular momentum distribution is then given by a summation over the 
radial modes:
\begin{equation}
n(j) = \sum_n \langle\hat\Psi_{jn} \hat\Psi^*_{jn}\rangle \, . 
\end{equation}

The angular momentum distribution for individual trajectories shows large occupation in particular angular modes, different for each run.  This indicates that vortices with different momenta appear each time.  For example, in one run, a vortex with $j=-1$ appears at about one quarter of the 
way 
through  the simulation, and persists until the end. The maximum occupation of the vortex
is around $n(j)=20$, owing to relatively small initial atom
numbers in this 2D trap simulation.  Shown in Fig. (3) is the ensemble average of the angular momentum distribution, which reveals quite a broad range of final angular momentum.  This is consistent with the existence of vortices.

In summary, we have demonstrated a three-dimensional
 real-time quantum dynamical simulation of Bose-condensation
with mesoscopic numbers of interacting atoms on a large lattice. 
Sampling errors and  lattice size restrictions impose strong
limitations on these initial  simulations.
The results, as well as
showing evidence for highly non-classical behaviour in a  first principles
simulation of  BEC formation, indicate the possibility of
spontaneous vortex formation in small evaporatively cooled condensates.

\section{Acknowledgement}
{ 
 This research was supported in part by the Australian Research
Council, and by the National Science Foundation under Grant No.
PHY94-07194.}

\begin{figure}
\caption{
Simulation of a two-dimensional Bose condensate,
showing the ensemble average (55 paths) atom density $ \langle n(k) \rangle$ along one
dimension in fourier space versus time.  }
\end{figure} 

\begin{figure}
\caption{
Simulation of a three-dimensional Bose condensate,
showing the ensemble average evolution (15 paths) of the confinement parameter $Q$.  }
\end{figure} 

\begin{figure}
\caption{
Ensemble average of the angular momentum distribution $\left< n(j)\right>$, during the condensation of a
two-dimensional Bose-condensate (40 paths).}
\end{figure}

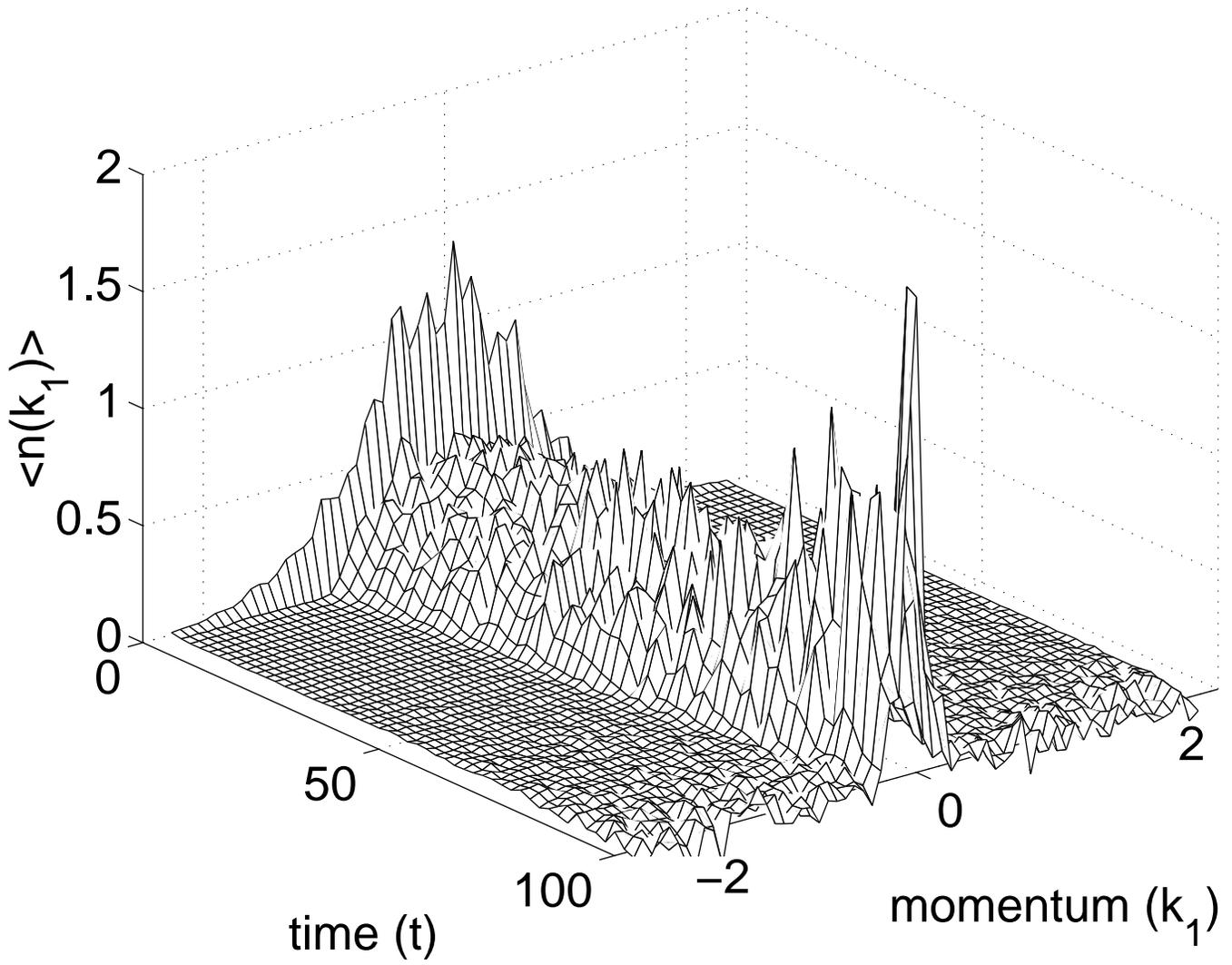
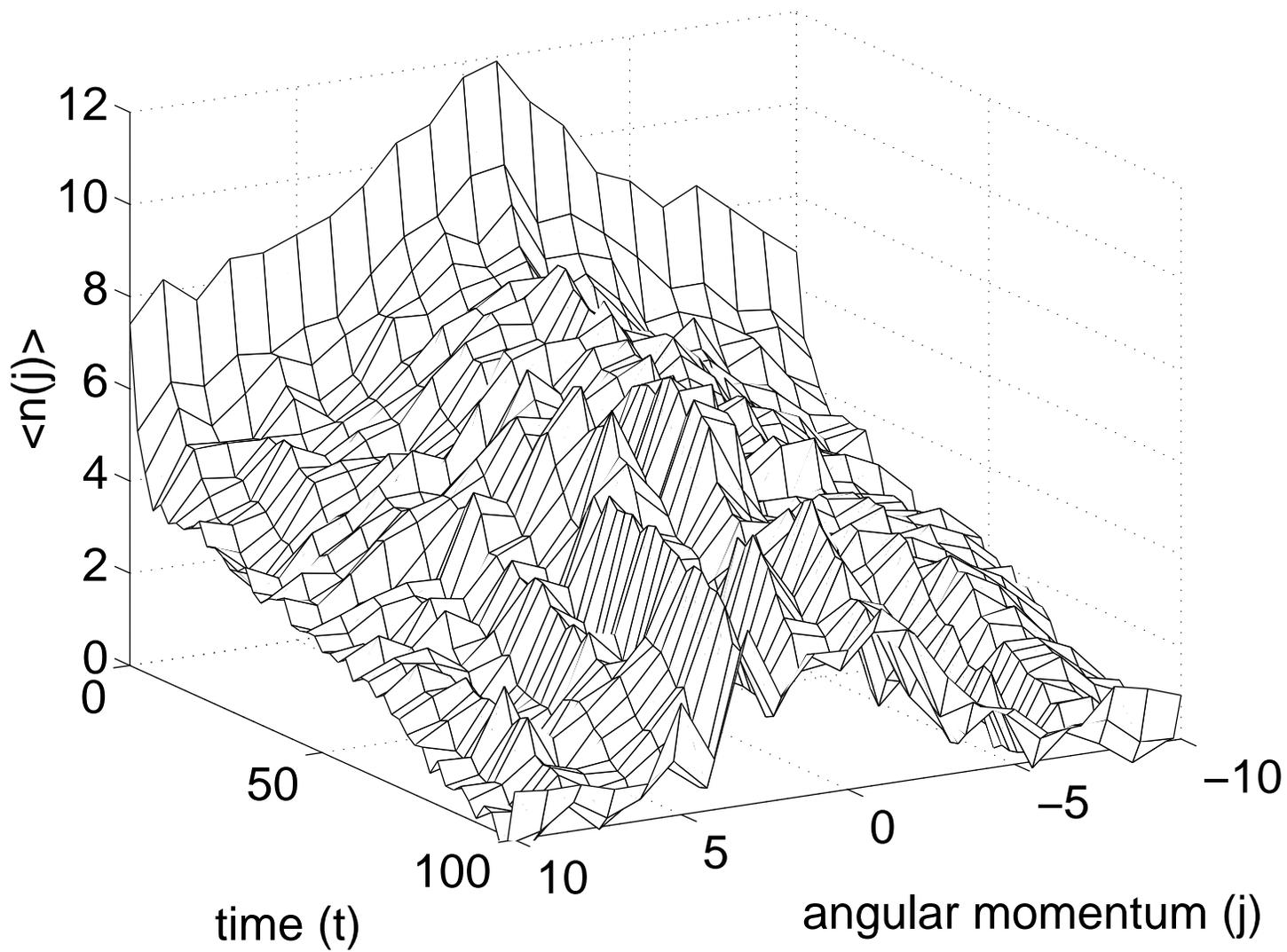
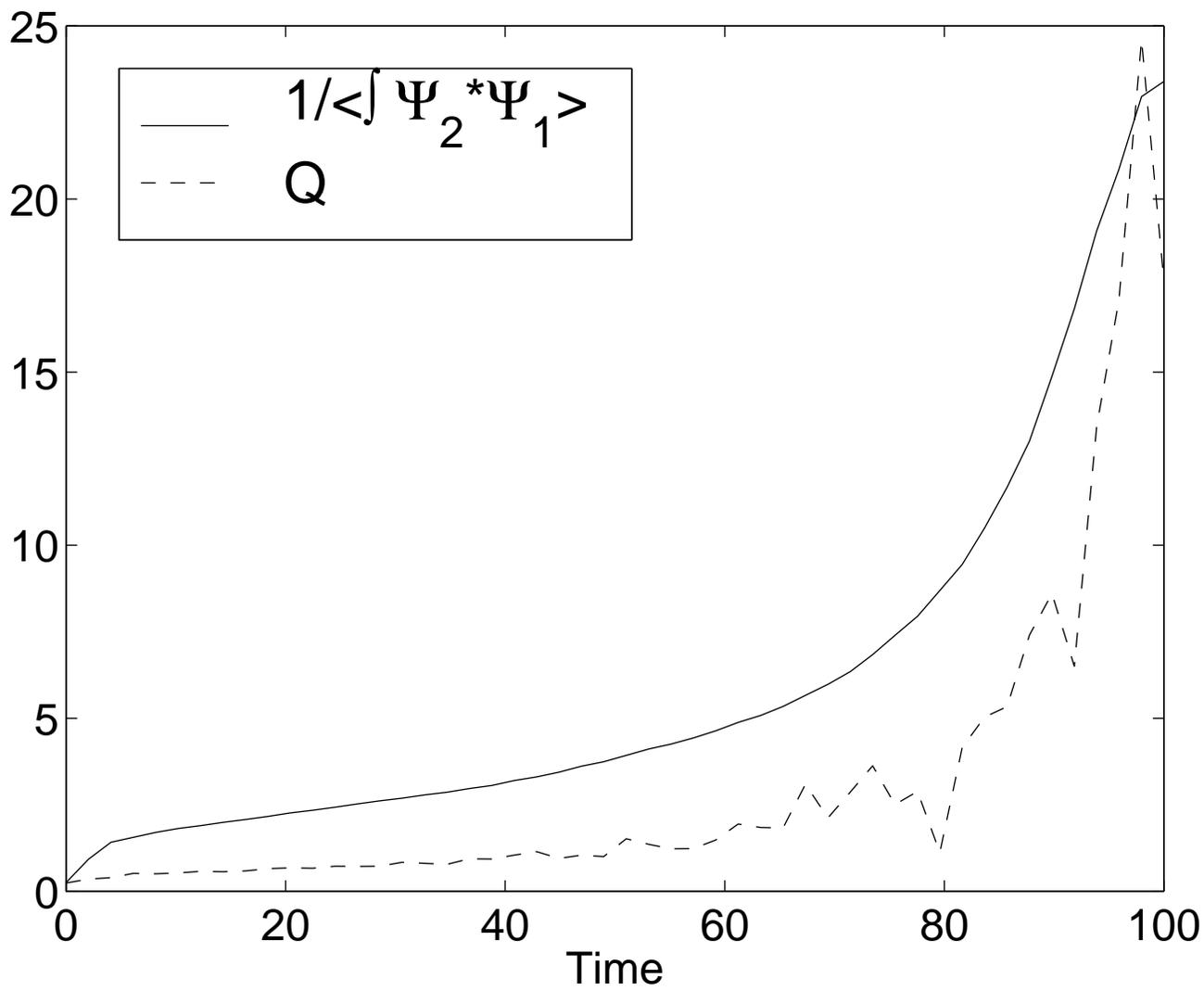
\end{document}